\documentclass[aps,prl,twocolumn,showpacs,superscriptaddress,preprintnumbers]{revtex4}
\usepackage{graphicx}
\usepackage{amsmath}
\def\vermark{}

\def\bsjetaBF{\ensuremath{(3.32\pm 0.87(\mathrm{stat.})^{+0.32}_{-0.28}(\mathrm{syst.})
\pm 0.42(f_s))\times 10^{-4}}}
\def\bsjetapBF{\ensuremath{(3.1\pm 1.2(\mathrm{stat.})^{+0.5}_{-0.6}(\mathrm{syst.})
\pm 0.38(f_s))\times 10^{-4}}}

\begin{document}

\makeatletter
\newcommand{\ps@mytitle}{%
  \renewcommand{\@oddhead}{{\vermark}\hfil}%
}
\makeatother

\title{Observation of the decay $B_s^0\to J/\psi\eta$ and Evidence for
$B_s^0\to J/\psi\eta'$}

\affiliation{Budker Institute of Nuclear Physics, Novosibirsk}
\affiliation{Chiba University, Chiba}
\affiliation{University of Cincinnati, Cincinnati, Ohio 45221}
\affiliation{Department of Physics, Fu Jen Catholic University, Taipei}
\affiliation{Justus-Liebig-Universit\"at Gie\ss{}en, Gie\ss{}en}
\affiliation{The Graduate University for Advanced Studies, Hayama}
\affiliation{Gyeongsang National University, Chinju}
\affiliation{Hanyang University, Seoul}
\affiliation{University of Hawaii, Honolulu, Hawaii 96822}
\affiliation{High Energy Accelerator Research Organization (KEK), Tsukuba}
\affiliation{Hiroshima Institute of Technology, Hiroshima}
\affiliation{University of Illinois at Urbana-Champaign, Urbana, Illinois 61801}
\affiliation{Indian Institute of Technology Guwahati, Guwahati}
\affiliation{Institute of High Energy Physics, Chinese Academy of Sciences, Beijing}
\affiliation{Institute of High Energy Physics, Vienna}
\affiliation{Institute of High Energy Physics, Protvino}
\affiliation{Institute of Mathematical Sciences, Chennai}
\affiliation{INFN - Sezione di Torino, Torino}
\affiliation{Institute for Theoretical and Experimental Physics, Moscow}
\affiliation{J. Stefan Institute, Ljubljana}
\affiliation{Kanagawa University, Yokohama}
\affiliation{Institut f\"ur Experimentelle Kernphysik, Universit\"at Karlsruhe, Karlsruhe}
\affiliation{Korea University, Seoul}
\affiliation{Kyoto University, Kyoto}
\affiliation{Kyungpook National University, Taegu}
\affiliation{\'Ecole Polytechnique F\'ed\'erale de Lausanne (EPFL), Lausanne}
\affiliation{Faculty of Mathematics and Physics, University of Ljubljana, Ljubljana}
\affiliation{University of Maribor, Maribor}
\affiliation{Max-Planck-Institut f\"ur Physik, M\"unchen}
\affiliation{University of Melbourne, School of Physics, Victoria 3010}
\affiliation{Nagoya University, Nagoya}
\affiliation{Nara University of Education, Nara}
\affiliation{Nara Women's University, Nara}
\affiliation{National Central University, Chung-li}
\affiliation{National United University, Miao Li}
\affiliation{Department of Physics, National Taiwan University, Taipei}
\affiliation{H. Niewodniczanski Institute of Nuclear Physics, Krakow}
\affiliation{Nippon Dental University, Niigata}
\affiliation{Niigata University, Niigata}
\affiliation{University of Nova Gorica, Nova Gorica}
\affiliation{Novosibirsk State University, Novosibirsk}
\affiliation{Osaka City University, Osaka}
\affiliation{Osaka University, Osaka}
\affiliation{Panjab University, Chandigarh}
\affiliation{Peking University, Beijing}
\affiliation{Princeton University, Princeton, New Jersey 08544}
\affiliation{RIKEN BNL Research Center, Upton, New York 11973}
\affiliation{Saga University, Saga}
\affiliation{University of Science and Technology of China, Hefei}
\affiliation{Seoul National University, Seoul}
\affiliation{Shinshu University, Nagano}
\affiliation{Sungkyunkwan University, Suwon}
\affiliation{School of Physics, University of Sydney, NSW 2006}
\affiliation{Tata Institute of Fundamental Research, Mumbai}
\affiliation{Excellence Cluster Universe, Technische Universit\"at M\"unchen, Garching}
\affiliation{Toho University, Funabashi}
\affiliation{Tohoku Gakuin University, Tagajo}
\affiliation{Tohoku University, Sendai}
\affiliation{Department of Physics, University of Tokyo, Tokyo}
\affiliation{Tokyo Institute of Technology, Tokyo}
\affiliation{Tokyo Metropolitan University, Tokyo}
\affiliation{Tokyo University of Agriculture and Technology, Tokyo}
\affiliation{Toyama National College of Maritime Technology, Toyama}
\affiliation{IPNAS, Virginia Polytechnic Institute and State University, Blacksburg, Virginia 24061}
\affiliation{Yonsei University, Seoul}
  \author{I.~Adachi}\affiliation{High Energy Accelerator Research Organization (KEK), Tsukuba} 
  \author{H.~Aihara}\affiliation{Department of Physics, University of Tokyo, Tokyo} 
  \author{K.~Arinstein}\affiliation{Budker Institute of Nuclear Physics, Novosibirsk}\affiliation{Novosibirsk State University, Novosibirsk} 
  \author{T.~Aso}\affiliation{Toyama National College of Maritime Technology, Toyama} 
  \author{V.~Aulchenko}\affiliation{Budker Institute of Nuclear Physics, Novosibirsk}\affiliation{Novosibirsk State University, Novosibirsk} 
  \author{T.~Aushev}\affiliation{\'Ecole Polytechnique F\'ed\'erale de Lausanne (EPFL), Lausanne}\affiliation{Institute for Theoretical and Experimental Physics, Moscow} 
  \author{T.~Aziz}\affiliation{Tata Institute of Fundamental Research, Mumbai} 
  \author{S.~Bahinipati}\affiliation{University of Cincinnati, Cincinnati, Ohio 45221} 
  \author{A.~M.~Bakich}\affiliation{School of Physics, University of Sydney, NSW 2006} 
  \author{V.~Balagura}\affiliation{Institute for Theoretical and Experimental Physics, Moscow} 
% ...
\collaboration{The Belle Collaboration}


\begin{abstract}
We report the first observation of $B_s^0\to J/\psi\eta$ and evidence
for $B_s^0\to J/\psi\eta'$.  These results are obtained from
$23.6\;\mathrm{fb}^{-1}$
of data collected at the $\Upsilon(5S)$ resonance with the Belle detector at
the KEKB $e^+e^-$ collider.  We measure the branching fractions
$\mathcal{B}(B_s^0\to J/\psi\eta)=\bsjetaBF$ with a significance of $7.3 \sigma$, and
$\mathcal{B}(B_s^0\to J/\psi\eta')=\bsjetapBF$ with a significance of $3.8 \sigma$.
\end{abstract}

\maketitle
\thispagestyle{mytitle}
\markright{\vermark}

\begin{figure}
\includegraphics[width=\columnwidth]
{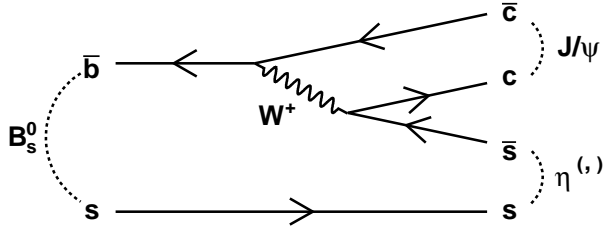}
\caption{\label{fig:bs_fyman}
Dominant diagram for the processes $B_s^0\to J/\psi\eta^{(\prime)}$.}
\end{figure}

The decays $B_s^0\to J/\psi \eta^{(\prime)}$ are dominated by the
$b\to c\bar c s$ process, as shown in Figure~\ref{fig:bs_fyman}.  The
$J/\psi \eta^{(\prime)}$ final states 
are $CP$ eigenstates, whose time distribution can be used
to measure directly the $B_s^0$ width difference
$\Delta\Gamma_s$~\cite{Dunietz:2000cr}.  Using SU(3) flavor symmetry,
an estimate of the $B_s^0\to J/\psi \eta^{(\prime)}$
branching fractions can be obtained relative to the
decay $B_d^0\to J/\psi K^0$~\cite{Skands:2000ru,Thomas:2007uy}:
\begin{equation*}
\frac{
\mathcal{B}(B_s^0\to J/\psi\eta^{(\prime)})}
{\mathcal{B}(B_d^0\to J/\psi K^0)} =
\sin^2(\cos^2)\phi_P\times p_{B_s^0}^{*3}/p_{B_d^0}^{*3},
\end{equation*}
where $p^*$ is the momentum of
$J/\psi$ or $\eta^{(\prime)}$ in the rest frame of the $B_s^0$ or
$B_d^0$.  Here, $\phi_P\simeq 37^\circ$ is the quark mixing angle in
the flavor basis with $\eta = \frac{1}{\sqrt{2}}(u\bar u+d \bar
d)\cos\phi_P - s\bar s\sin\phi_P$.  We then expect $\mathcal{B}(B_s^0\to
J/\psi\eta^{(\prime)})\sim 3.45(4.88)\times 10^{-4}$, using the value
$\mathcal{B}(B_d^0\to J/\psi K^0) = 8.71\times
10^{-4}$~\cite{Amsler:2008zzb}.  To date there is only the upper limit
$\mathcal{B}(B_s^0\to J/\psi\eta) < 3.8\times
10^{-3}$ at $90\%$ confidence level~\cite{Acciarri:1996ur}.  Recently,
a large sample of $B_s^0$ mesons was produced from $e^+e^-$ collisions
at the $\Upsilon(5S)$ resonance, where final states with photons can
be reconstructed with low background.

In this letter, we report measurements of fully reconstructed
$B_s^0\to J/\psi\eta$ and $B_s^0\to J/\psi\eta'$ decays using a
$(23.6\pm 0.3)\;\mathrm{fb}^{-1}$ data sample collected with the Belle
detector at the KEKB asymmetric-energy $e^+e^-$ collider~\cite{KEKB}
operated at the $\Upsilon(5S)$ resonance.  The beam energy in the
center-of-mass (CM) frame $E_\mathrm{beam}$ is measured to be
$5433.5\pm 0.5$ MeV using $\Upsilon(5S)\to\Upsilon(1S)\pi^+\pi^-$,
$\Upsilon(1S)\to\mu^+\mu^-$ decays~\cite{Abe:2007tk}.  The total
$b\bar b$ cross section at the $\Upsilon(5S)$ energy has been measured
to be $\sigma_{b\bar b}^{\Upsilon(5S)} = (0.302\pm
0.014)\;\mathrm{nb}$~\cite{Drutskoy:2006fg}.  $B_s^0$ events are
produced from $B_s^{(*)}\bar B_s^{(*)}$ pairs where three $B_s^0$
production channels $B_s^0 \bar B_s^0,\;B_s^*\bar B_s^0,\;B_s^*\bar
B_s^*$ are kinematically allowed.  The total fraction of
$B_s^{(*)}\bar B_s^{(*)}$ pairs in $b\bar b$ events has been measured
to be $f_s = N_{B_s^{(*)}\bar B_s^{(*)}}/N_{b\bar b}
=(19.5^{+3.0}_{-2.3})\%$~\cite{Drutskoy:2006fg}.  Thus, the
$23.6\;\mathrm{fb}^{-1}$ data sample contains a total of $2.78$ million
$B_s^0$ mesons.  The fractions of $B_s^0$ production channels are measured to be
$f_{B_s^*\bar B_s^*}=N_{B_s^*\bar B_s^*}/N_{B_s^{(*)}\bar B_s^{(*)}}
=(90.1^{+3.8}_{-4.0}\pm 0.2)\%$, $f_{B_s^*\bar B_s^0}=N_{B_s^*\bar
B_s^0}/N_{B_s^{(*)}\bar B_s^{(*)}} =(7.3^{+0.33}_{-0.30}\pm
0.1)\%$~\cite{:2008sc}.

The Belle detector is a large-solid-angle magnetic spectrometer
that consists of a silicon vertex detector (SVD), a 50-layer central drift
chamber (CDC), an array of aerogel threshold Cherenkov counters (ACC), a
barrel-like arrangement of time-of-flight scintillation counters (TOF), and
an electromagnetic calorimeter (ECL) comprised of CsI(Tl) crystals
located inside a superconducting solenoid coil that provides a 1.5 T magnetic field.
An iron flux-return located outside the coil is instrumented to detect
$K_L^0$ mesons and identify muons (KLM).  The detector is described in detail
elsewhere~\cite{:2000cg}.

Charged tracks are required to originate within $0.5$ cm in the
radial direction and within 5 cm in the beam direction,
with respect to the interaction point.  Electron candidates
are identified by combining information from the ECL, the CDC $(dE/dx)$, and
the ACC.  Muon candidates are identified through track penetration depth
and hit patterns in the KLM system.  Identification of pions is based on
combining information from the CDC $(dE/dx)$, the TOF and the ACC.

Two oppositely charged leptons $l^+l^-$ ($l=e\;\mathrm{or}\;\mu$) and
bremsstrahlung photons lying within 50 mrad of $e^+$ or $e^-$ tracks
are combined to form a $J/\psi$ meson.  The leptons are
required to be positively identified as electrons or muons.  The
invariant mass is required to lie in the ranges
$-0.150\;\mathrm{GeV}/c^2<M_{ee(\gamma)}-
m_{J/\psi}<0.036\;\mathrm{GeV}/c^2$ and
$-0.060\;\mathrm{GeV}/c^2<M_{\mu\mu}- m_{J/\psi}<0.036\;\mathrm{GeV}/c^2$,
where $m_{J/\psi}$ denotes the nominal $J/\psi$ mass, and
$M_{ee(\gamma)}$ and $M_{\mu\mu}$ are the reconstructed invariant masses
for $e^+e^-(\gamma)$ and $\mu^+\mu^-$, respectively.

Photon candidates are selected from ECL showers not associated with
charged tracks.  An energy deposition with a photon-like shape and an energy greater
than 50 MeV is required.  $\pi^0$ candidates are selected by combining two photon
candidates with an invariant mass in the range
$115\;\mathrm{MeV}/c^2<M_{\gamma\gamma} <155\;\mathrm{MeV}/c^2$ .

Candidate $\eta$ mesons are reconstructed in the $\gamma\gamma$ and
$\pi^+\pi^-\pi^0$ final states.  We require the invariant mass to be
in the range $500\;\mathrm{MeV}/c^2 < M_{\gamma\gamma} <
575\;\mathrm{MeV}/c^2\;([-3.5\sigma,2.0\sigma])$ and
$535\;\mathrm{MeV}/c^2 < M_{\pi^+\pi^-\pi^0} <
560\;\mathrm{MeV}/c^2\;(\pm 2.5\sigma)$.  In the $\gamma\gamma$
final state, we require that $|\cos\theta_\mathrm{dec}|<0.9$, to
reduce background from combinatorial $\gamma$'s, where $\theta_\mathrm{dec}$ is the
angle between the $\gamma$ and $\eta$ lab momentum direction in the
$\eta$ rest frame.

Candidate $\eta'$ mesons are reconstructed in the $\eta\pi^+\pi^-$ and
$\rho^0\gamma$ channels.  The $\eta$ candidates are selected in the
same two channels as above.  Thus, there are three sub-channels for
$\eta'$ reconstruction.  $\rho^0$'s are selected from oppositely charged
pion pairs  satisfying $550 \;
\mathrm{MeV}/c^2<M_{\pi^+\pi^-}<900\;\mathrm{MeV}/c^2$ and a helicity
requirement $|\cos\theta_\mathrm{hel}|<0.85$ since the $\rho^0$ is
longitudinally polarized.  Here $\theta_\mathrm{hel}$ is the helicity
angle of $\rho^0$, calculated as the angle between the direction of
the $\pi^+$ and the direction opposite to the $\eta'$ momentum in the
$\rho^0$ rest frame.  We require the reconstructed $\eta'$ invariant
mass to satisfy $ 940\;\mathrm{MeV}/c^2< M_{\eta'} <
975\;\mathrm{MeV}/c^2\;(\pm 3\sigma)$.

We combine $J/\psi$ and $\eta^{(\prime)}$ candidates to form $B_s^0$ mesons.
Signal candidates are identified by two kinematic variables computed
in the $\Upsilon(5S)$ frame: the energy difference $\Delta E=
E_B^*-E_\mathrm{beam}$ and the beam-energy constrained mass
$M_\mathrm{bc}=\sqrt{(E_\mathrm{beam})^2- (p_B^*)^2}$, where $E_B^*$
and $p_B^*$ are the energy and momentum of the reconstructed $B_s^0$
candidate.
To improve the $\Delta E$ and $M_\mathrm{bc}$ resolutions, mass-constrained
kinematic fits are applied to $J/\psi$, $\eta^{(\prime)}$ and $\pi^0$
candidates.  We retain $B_s^0$ meson candidates with $|\Delta E|<0.4$ GeV and
$M_\mathrm{bc}>5.25\;\mathrm{GeV}/c^2$ for further analysis.
If there are multiple candidates in a single
event, we choose the candidate that minimizes the sum of the $\chi^2$'s of the
mass-constrained fits.

To suppress the two-jet-like continuum background from $e^+e^-\to
q\bar q\;(q=u,d,s,c)$, we require the ratio of second to zeroth
Fox-Wolfram moments~\cite{Fox:1978vu} to be less than $0.4$.  This
requirement is optimized by maximizing a figure-of-merit
$N_\mathrm{S}/\sqrt{N_\mathrm{S}+N_\mathrm{B}}$, where $N_S$ is the
expected number of signal events and $N_B$ is the number of continuum
events, in the $B^*\bar B^*$ signal region.  The continuum background
is modeled by a second-order polynomial in $\Delta E$ and an ARGUS
function~\cite{Albrecht:1986nr} in $M_\mathrm{bc}$.  We study the
continuum background in a $J/\psi$ sideband defined as
$2.5\;\mathrm{GeV}/c^2 <M_{J/\psi}<3.4\;\mathrm{GeV}/c^2$, excluding
the region $-0.200(-0.080)\;\mathrm{GeV}/c^2<M_{J/\psi}- m_{J/\psi}
<0.048\;\mathrm{GeV}/c^2$ for $J/\psi\to e^+e^-(\mu^+\mu^-)$ in data.
The shapes for continuum backgrounds for each sub-channel are
determined from fits to $J/\psi$ sideband data with a relaxed lepton
identification requirement.  The yields for continuum background are
determined from the yields in $J/\psi$ sideband data with all
selections other than the $J/\psi$ mass cut as in the nominal
selection.  Scale factors are determined from the ratio of the
$J/\psi$ selection area to the $J/\psi$ sideband area by fitting Monte
Carlo (MC) $J/\psi$ mass distributions.  For continuum $J/\psi$ backgrounds that
may not be well modeled in MC, we studied the off-resonance data and
obtained the fractions of real $J/\psi$'s in continuum, which are
applied as a correction to the continuum background yields.

The remaining background is from $B\bar B$ ($B=B_s^0,\;B_d^0,\;B_u^\pm$)
events with one $B$ meson decays to a final state with $J/\psi$ (denoted
$J/\psi X$).  We use a MC sample generated at the $\Upsilon(5S)$
resonance that includes all known $B\to J/\psi X$
processes to estimate this background, 
and find that the dominant contribution is from
$B_d^0,B_u^\pm\to J/\psi +\mathrm{strange\ mesons}$. 
This background does not peak in either $\Delta
E$ and $M_\mathrm{bc}$ and is described by an exponential function in
$\Delta E$ and an ARGUS function in $M_\mathrm{bc}$, with shapes
determined from MC.  For the channel $J/\psi \eta'(\rho^0\gamma)$, the
size of the $J/\psi X$ background is comparable to the expected signal
yield in the $B_s^*\bar B_s^*$ signal region, so we use a $\eta'$ sideband
defined as $0.90\;\mathrm{GeV}/c^2<M_{\eta'}<0.93\;\mathrm{GeV}/c^2$
and $0.99\;\mathrm{GeV}/c^2<M_{\eta'}<1.02\;\mathrm{GeV}/c^2$ to study
it.  We fit the $\eta'$ sideband data to obtain the $J/\psi X$ background
shape for the $J/\psi \eta'(\rho^0\gamma)$ channel,
where the continuum contribution is fixed from the $J/\psi$ sideband.



We categorize events into two $\eta$ and three $\eta'$ sub-channels.
For the $J/\psi\eta$ and $J/\psi\eta'$ modes, we perform a
simultaneous unbinned maximum likelihood fit to the $\Delta
E-M_\mathrm{bc}$ distributions for each group of two $\eta$ or
three $\eta'$ sub-channels, with the signal branching fraction as a
common parameter while treating the signal and background shapes
separately for the different $\eta^{(\prime)}$ sub-channels.

\begin{figure}
\setlength{\unitlength}{0.005\columnwidth}
\includegraphics[width=0.5\columnwidth]
{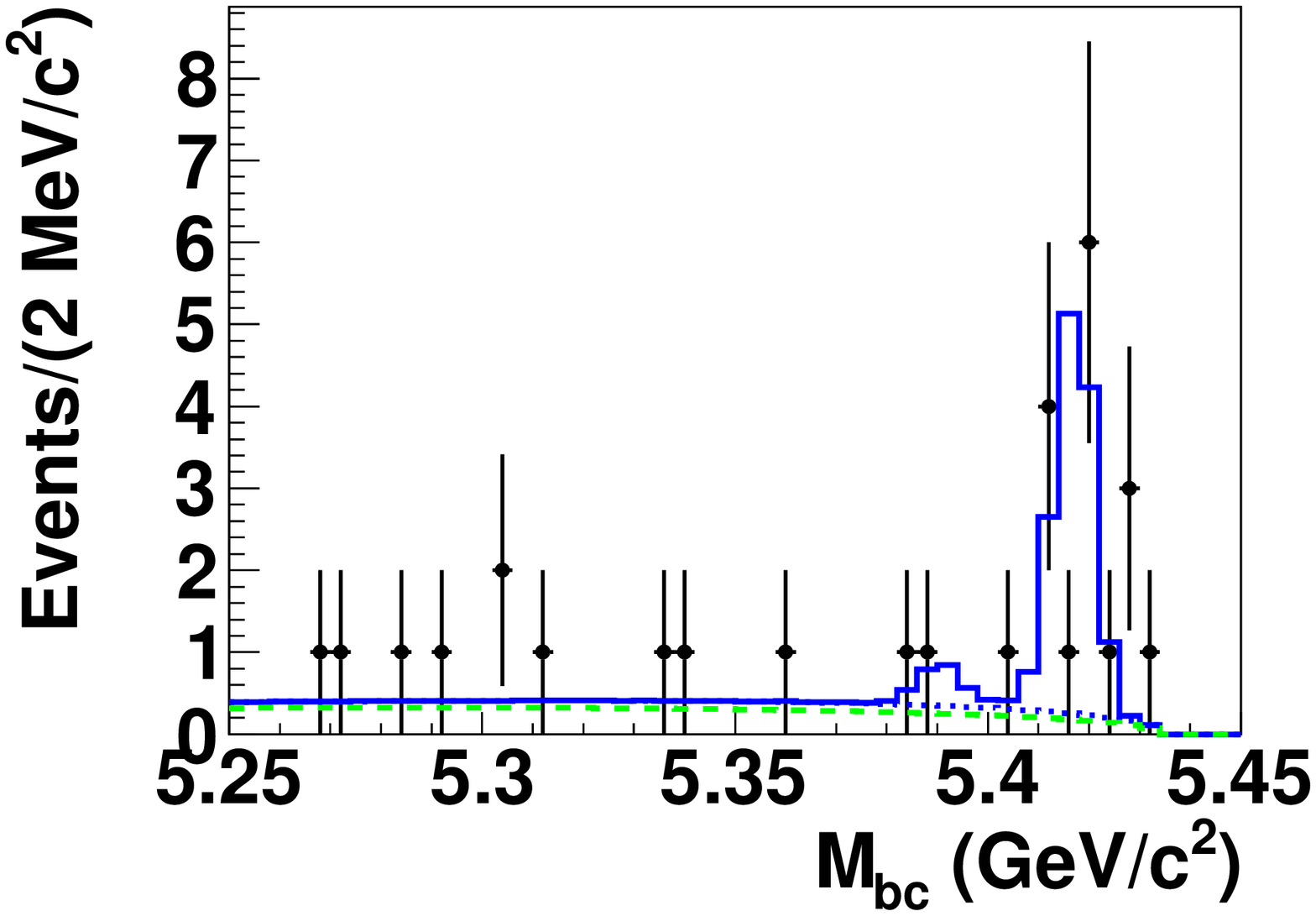}%
\includegraphics[width=0.5\columnwidth]
{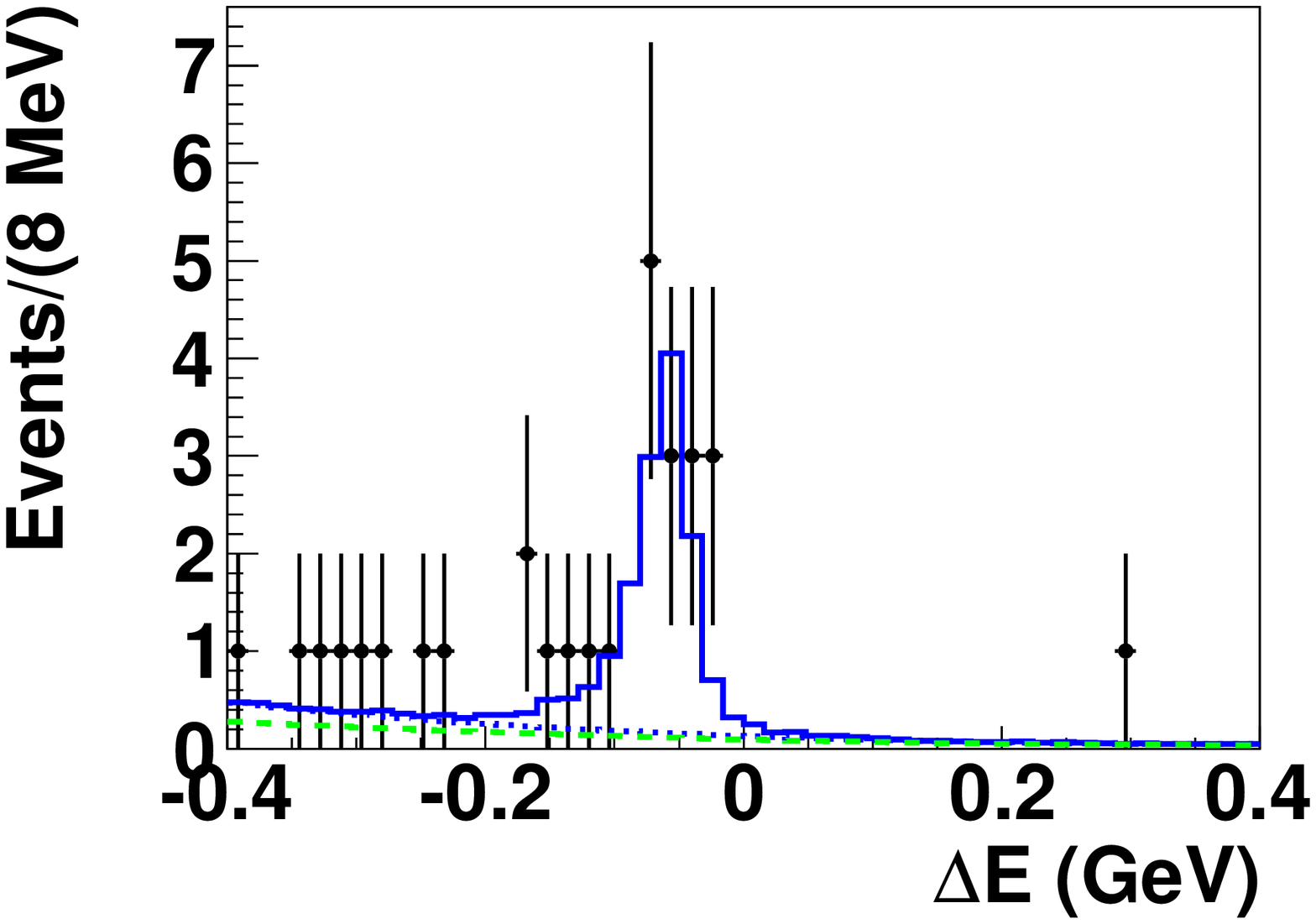}%
\hspace{-1\columnwidth}%
\begin{picture}(199,0)
\put(25,55){(a)}
\put(125,55){(b)}
\end{picture}\hfill%
\caption{\label{fig:5S_scan-eA-hde_mbc}
Distributions for the  $J/\psi\eta$ channels, (a) $M_\mathrm{bc}$ distribution
for the $B_s^*\bar B_s^*$ signal region with
$\Delta E \in [-97,-9]$ MeV,
and (b) $\Delta E$ distribution for the $B_s^*\bar B_s^*$ signal region
with $M_\mathrm{bc}\in[5.405,5.431]\;\mathrm{GeV}/c^2$.
The full histogram shows projections of fit results.
The sum of all backgrounds is represented by the blue dotted curves
with the green dashed curve corresponding to continuum background. }
\end{figure}

\begin{figure}
\setlength{\unitlength}{0.005\columnwidth}
\includegraphics[width=0.5\columnwidth]
{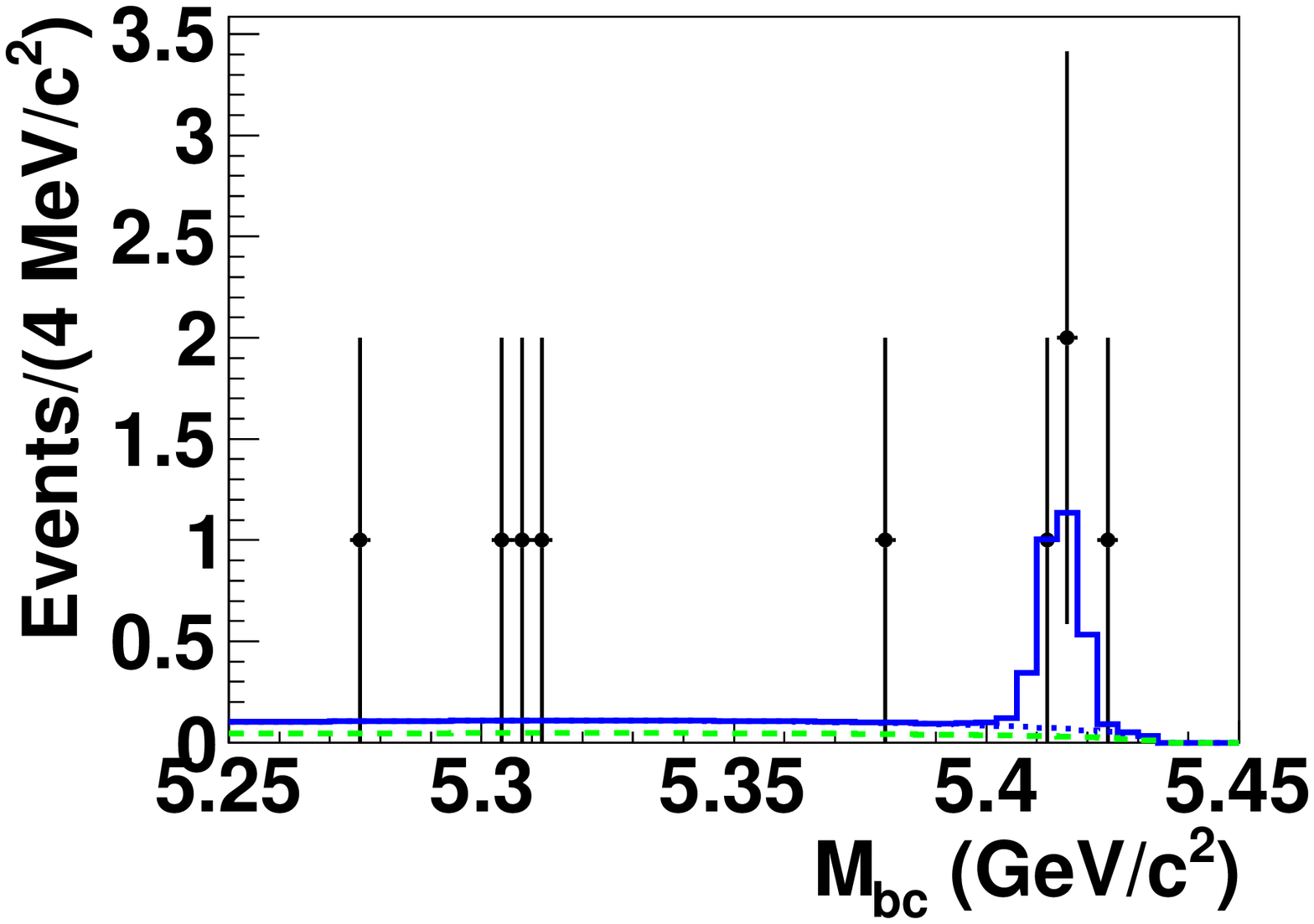}%
\includegraphics[width=0.5\columnwidth]
{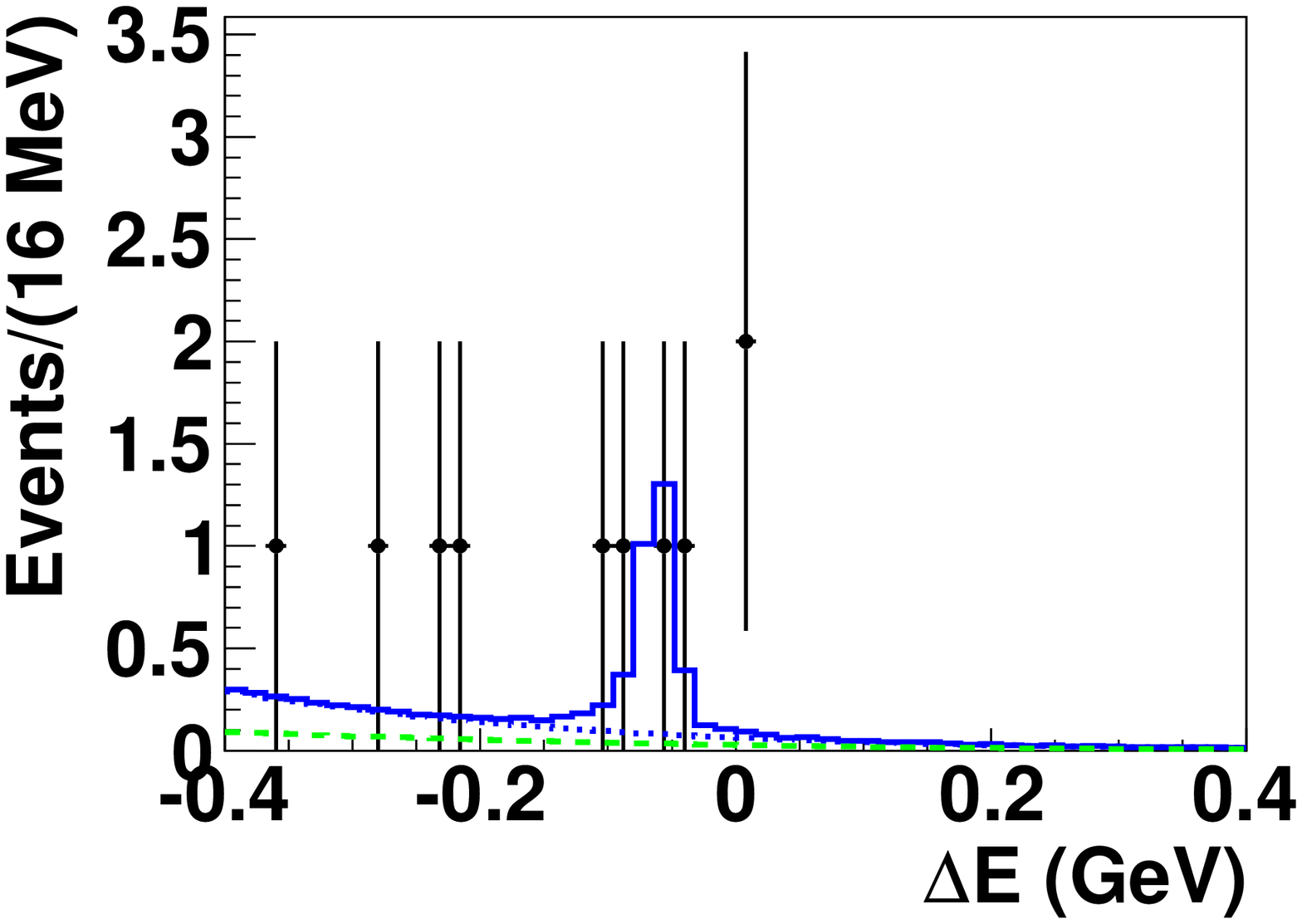}%
\hspace{-1\columnwidth}%
\begin{picture}(199,0)
\put(25,55){(a)}
\put(125,55){(b)}
\end{picture}\hfill\\
\includegraphics[width=0.5\columnwidth]
{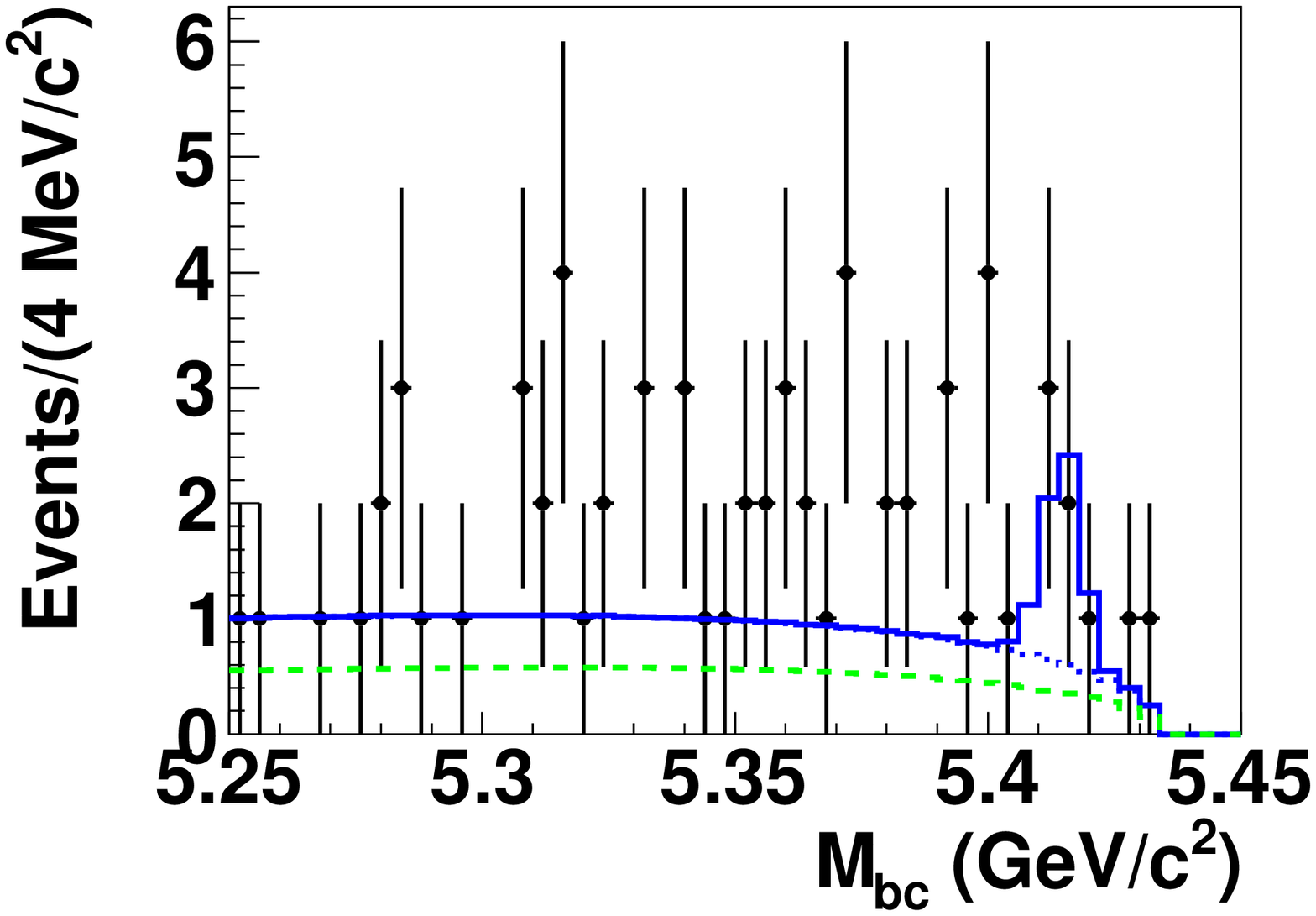}%
\includegraphics[width=0.5\columnwidth]
{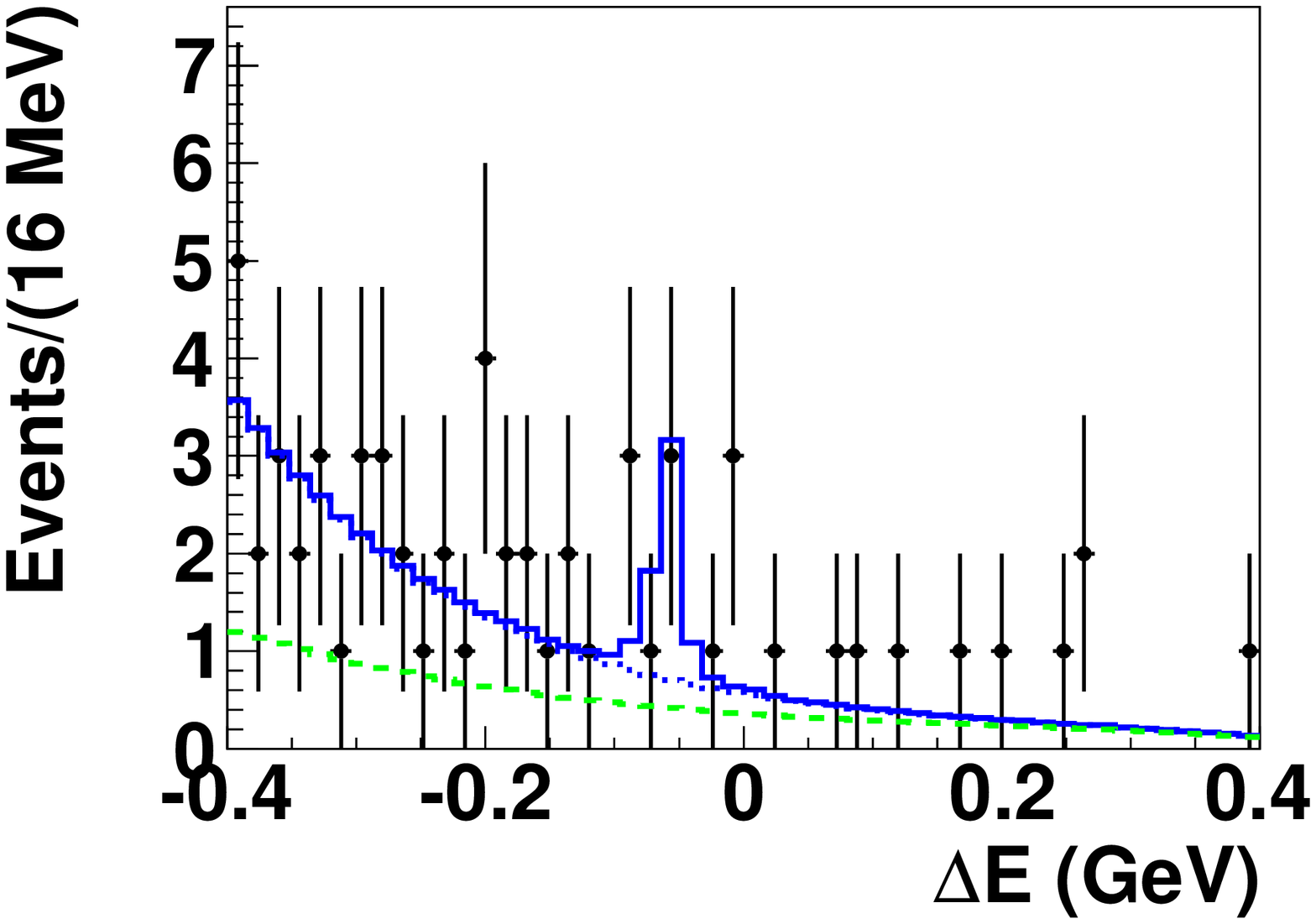}%
\hspace{-1\columnwidth}%
\begin{picture}(199,0)
\put(25,55){(c)}
\put(125,55){(d)}
\end{picture}\hfill\\
\caption{\label{fig:5S_scan-epA-hmbc}
Fit projections for the combined 
clean $J/\psi\eta'(\eta\pi^+\pi^-)$ channels (a,b) and 
$J/\psi\eta'(\rho^0\gamma)$ channel (c,d). 
The projections are shown in the $B_s^*\bar B_s^*$ signal region with
$\Delta E\in [-98,-17]$ MeV (a,c), and with
$M_\mathrm{bc}\in [5.403,5.428]\;\mathrm{GeV}/c^2$ (b,d).
The sum of all backgrounds is represented by the blue dotted curves
with the green dashed curve corresponding to continuum background.}
\end{figure}

\begin{table*}
\caption{\label{tab:jAeA-N}
A summary of efficiencies, significances
and branching fractions ($\mathcal{B}$) for each mode.
} 
\begin{ruledtabular}
\begin{tabular}{lccc}
mode & $\mathcal{B}_i\epsilon_i$  & Significance & $\mathcal{B}$\\
\hline
$B_s^0\to J/\psi\eta(\gamma\gamma)$     & 1.34\%   &  & \\
$B_s^0\to J/\psi\eta(\pi^+\pi^-\pi^0 )$ & 0.45\%   &  & \\
Total $B_s^0\to J/\psi\eta$    & 1.79\% &  $7.3\sigma$&
\bsjetaBF \\
\hline
$B_s^0\to J/\psi\eta'(\eta\pi\pi)$     & 0.52\%   &  &\\
$B_s^0\to J/\psi\eta'(\rho^0\gamma)$   & 0.88\%   &  &\\
Total $B_s^0\to J/\psi\eta'$ & 1.40\% &  $3.8 \sigma$&
\bsjetapBF
\end{tabular}
\end{ruledtabular}
\end{table*}

\begin{table}
\caption{\label{tab:errBF_syst}
Relative systematic errors (in \%) for $\mathcal{B}(J/\psi\eta^{(\prime)})$.}
\begin{ruledtabular}
\begin{tabular}{lcc}
Source  &  $\mathcal{B}(J/\psi\eta)$ & $\mathcal{B}(J/\psi\eta')$\\ \hline
Signal shape calibration &  $+5.8, - 2.9$ &  $+11.7, -16.8$ \\
Beam energy &  $+1.6, - 0.0$ &   $+4.8, -4.3 $  \\
MC signal shape &  $+1.0, - 2.0$ & $+2.6, - 4.0$  \\
$f_{B_s^{(*)}\bar B_s^{(*)}}$  & $+0.7,-1.5$  &  $+4.6, -4.0$ \\
Background parameters & $+0.9,-0.8$ & $+6.0,-5.5$ \\  \hline
Track reconstruction  & 2.5 & 4.2  \\
Lepton identification & 4.2 & 4.1\\
Pion identification   & 0.4  &  2.3  \\
$\eta(\pi^0)\to\gamma\gamma$ selection&  4.1 & 2.8  \\
$\mathcal{B}(J/\psi \to ll)$ & 0.72 & 0.72 \\
$\mathcal{B}(\eta^{(\prime)}\to \mathrm{final\ states})$ &
   0.49 & 2.3 \\
Luminosity & 1.3 & 1.3 \\
$\sigma_{b\bar b}$ &  4.6 & 4.6 \\
$f_s$  & $+13.4,-13.3$ & $+13.4,-13.3$ \\
\hline
Total  & $+16.8,-16.1$ & $+21.9, -24.8$\\
\end{tabular}
\end{ruledtabular}
\end{table}

In the fit, the signal normalization for each $B_s^0$ production channel
is parameterized as $N_\mathrm{sig} = 2\times
N_{B_s^{(*)}\bar B_s^{(*)}} f_{B_s^{(*)}\bar B_s^{(*)}}
\mathcal{B}(B_s^0\to J/\psi\eta^{(\prime)})
\Sigma_i\mathcal{B}_i \epsilon_i $.
We use three fractions $f_{B_s^*\bar B_s^*}$, $f_{B_s^*\bar B_s^0}$,
and $f_{B_s^0\bar B_s^0} = 1 - f_{B_s^*\bar B_s^*} - f_{B_s^*\bar
B_s^0}$.  In the $J/\psi \eta^{\vphantom{(}\prime}$ mode,
because of low statistics, we only
include the $B_s^*\bar B_s^*$ channel in the fit.
The index $i$ denotes each $\eta^{(\prime)}$ sub-channel.
The product 
$\mathcal{B}_i= \mathcal{B}(J/\psi\to l^+l^-)\mathcal{B}_i(\eta^{(\prime)})$ is
the total branching fraction to final states with a $J/\psi$ and an $\eta^{(\prime)}$,
and $\epsilon_i$ is the MC reconstruction efficiency.
The values of the weighted efficiencies
$\mathcal{B}_i\epsilon_i$ are listed in Table~\ref{tab:jAeA-N}.
The signal shapes are from signal MC histograms while the means and widths
of the distributions are
corrected using a $B^+\to J/\psi K^{*+}(K^{*+}\to K^+\pi^0)$
control sample from $\Upsilon(4S)$.
The continuum background's shapes and yields and the $J/\psi X$ background's
shapes are fixed.
The floating parameters for each fit are the branching fraction
$\mathcal{B}(B_s^0\to J/\psi\eta^{(\prime)})$ and the $J/\psi X$ background yields
for each sub-channel.

The projections of the fit in the $B_s^*\bar B_s^*$ signal region are
shown in Figures~\ref{fig:5S_scan-eA-hde_mbc} and
\ref{fig:5S_scan-epA-hmbc}.  The signal efficiencies, branching fractions,
and significances including systematic uncertainties are listed in
Table~\ref{tab:jAeA-N}.  We calculate a total of $14.9\pm 4.1$
$B_s^0\to J/\psi\eta$ events and $10.7\pm 4.6$ $B_s^0\to J/\psi\eta'$ events
in the $\Upsilon(5S)\to B_s^{*}\bar B_s^{*}$ channel.

The $B_s^0\to J/\psi\eta$ decay is observed for the first time
and evidence for the  $B_s^0\to J/\psi\eta'$  decay is found.
The significance is defined by $S =
\sqrt{2\ln(\mathcal{L}_\mathrm{max}/\mathcal{L}_0)}$, where
$\mathcal{L}_\mathrm{max}(\mathcal{L}_0)$ is the likelihood value at
the maximum (with the signal branching fraction set to zero).  The
significance including systematic uncertainty is taken as the smallest
value for each systematic variation described below.

The muon and pion identification efficiencies from MC are calibrated
using the $J/\psi\to l^+l^-$ and $D^{*+}\to D^0\pi^+$ control samples
in data, respectively.  The total systematic error due to lepton
identification is weighted to be $4.2(4.1)\%$ for $J/\psi\eta^{(\prime)}$
modes.  The systematic error due to pion identification is
$0.4(2.3)\%$ for the $J/\psi\eta^{(\prime)}$ modes.

The systematic errors due to the signal shape mean and width
corrections and background parameters are determined
by varying each parameter by its error, repeating the fit,
and summing the shifts in branching fraction in quadrature.
The beam energy has an error of $\pm 0.5\;\mathrm{MeV}$,
whose systematic effect is evaluated by varying
the mean value of $\Delta E$ and $M_\mathrm{bc}$ for the signal shapes
simultaneously according to the uncertainty in the beam energy.
All the systematic errors are summarized in Table~\ref{tab:errBF_syst}. 
The large systematic errors due to $f_s$ are quoted separately in the final
results.

The ratio of the two branching fractions $R=\mathcal{B}(B_s^0\to
J/\psi\eta')/\mathcal{B}(B_s^0\to J/\psi\eta)$ is also calculated.
The statistical errors of the two modes are combined.  The common
systematic errors due to luminosity, cross-section and $f_s$ cancel.
Correlated systematic errors due to calibration, beam energy, track
reconstruction and particle identification are treated properly by
varying the numerator and denominator simultaneously.
Other systematic sources in the two branching fractions are treated independently.

In summary, we observe $B_s^0\to J/\psi\eta$ decay with a significance
of $7.3\sigma$ and find evidence for $B_s^0\to J/\psi\eta'$ with a
significance of $3.8\sigma$.  We measure the branching fractions
$\mathcal{B}(B_s^0\to J/\psi\eta)=\bsjetaBF$ and $\mathcal{B}(B_s^0\to
J/\psi\eta')=\bsjetapBF$.  The ratio of two branching fractions is
measured as $R=\frac{\mathcal{B}(B_s\to J/\psi
\eta)}{\mathcal{B}(B_s\to J/\psi \eta')} =
0.924^{+0.52}_{-0.44}(\mathrm{stat.}) ^{+0.15}_{-0.20}(\mathrm{syst.})
$.  The results are consistent with SU(3) expectations using the
measured value of $\mathcal{B}(B_d^0\to J/\psi
K^0)$~\cite{Skands:2000ru,Thomas:2007uy}.

We thank the KEKB group for excellent operation of the
accelerator, the KEK cryogenics group for efficient solenoid
operations, and the KEK computer group and
the NII for valuable computing and SINET3 network support.  
We acknowledge support from MEXT, JSPS and Nagoya's TLPRC (Japan);
ARC and DIISR (Australia); NSFC (China); 
DST (India); MEST, KOSEF, KRF (Korea); MNiSW (Poland); 
MES and RFAAE (Russia); ARRS (Slovenia); SNSF (Switzerland); 
NSC and MOE (Taiwan); and DOE (USA).


\begin{thebibliography}{99}
\bibitem{Dunietz:2000cr}
  I.~Dunietz, R.~Fleischer and U.~Nierste,
  Phys.\ Rev.\  D {\bf 63}, 114015 (2001)
  [arXiv:hep-ph/0012219].

\bibitem{Skands:2000ru}
  P.~Z.~Skands,
  JHEP {\bf 0101}, 008 (2001)
  [arXiv:hep-ph/0010115].

\bibitem{Thomas:2007uy}
  C.~E.~Thomas,
  JHEP {\bf 0710}, 026 (2007)
  [arXiv:0705.1500 [hep-ph]].

\bibitem{Amsler:2008zzb}
  C.~Amsler {\it et al.}  [Particle Data Group],
  Phys.\ Lett.\  B {\bf 667}, 1 (2008).

\bibitem{Acciarri:1996ur}
  M.~Acciarri {\it et al.}  [L3 Collaboration],
  Phys.\ Lett.\  B {\bf 391}, 481 (1997).

\bibitem{KEKB}
S.~Kurokawa and E.~Kikutani, Nucl. Instr. and. Meth. A
{\bf 499}, 1 (2003).

\bibitem{:2000cg}
  A.~Abashian {\it et al.} [Belle Collaboration],
  Nucl.\ Instrum.\ Meth.\  A {\bf 479}, 117 (2002).

\bibitem{Abe:2007tk}
  K.~F.~Chen {\it et al.}  [Belle Collaboration],
  Phys.\ Rev.\ Lett.\  {\bf 100}, 112001 (2008)
  [arXiv:0710.2577 [hep-ex]].

\bibitem{Drutskoy:2006fg}
  A.~Drutskoy {\it et al.}  [Belle Collaboration],
  Phys.\ Rev.\ Lett.\  {\bf 98}, 052001 (2007),
  G.~S.~Huang {\it et al.}  [CLEO Collaboration],
  Phys.\ Rev.\  D {\bf 75}, 012002 (2007).
We are using the $\sigma_{b\bar b}^{\Upsilon(5S)}$ value, $0.302\pm 0.014$ nb,
which is an average of these two measurements.

\bibitem{:2008sc}
  R.~Louvot {\it et al.}  [Belle Collaboration],
  Phys.\ Rev.\ Lett.\  {\bf 102}, 021801 (2009)
  [arXiv:0809.2526 [hep-ex]].

\bibitem{Fox:1978vu}
 The Fox-Wolfram moments were introduced in 
  G.~C.~Fox and S.~Wolfram,
  Phys.\ Rev.\ Lett.\  {\bf 41}, 1581 (1978).

\bibitem{Albrecht:1986nr}
  H.~Albrecht {\it et al.}  [ARGUS Collaboration],
  Phys.\ Lett.\  B {\bf 185}, 218 (1987).

\end{thebibliography}
\end{document}